\documentclass[lettersize,journal]{IEEEtran}
\usepackage{amsmath,amsfonts}
\usepackage{algorithm}
\usepackage{array}
\usepackage[caption=false,font=normalsize,labelfont=sf,textfont=sf]{subfig}
\usepackage{textcomp}
\usepackage{stfloats}
\usepackage{url}
\usepackage{verbatim}
\usepackage{graphicx}
\usepackage{cite}
\usepackage{xcolor}
\usepackage{hyperref}
\usepackage{graphicx}
\usepackage{float}

\usepackage{algpseudocode}

\begin{document}

\title{Complexity Assessment of Analog and Digital Security Primitives Signals Using the Disentropy of Autocorrelation}

\author{Paul~Jimenez, Raphael~Cardoso, Maurício~Gomes~de~Queiroz, Mohab~Abdalla, Cédric~Marchand, Xavier~Letartre, and Fabio~Pavanello

\thanks{This work was supported by the French Agence Nationale de la Recherche under project number ANR-20-CE39-0004 - PHASEPUF project. F.P., X.L., and C.M. acknowledge funding from the European Union’s Horizon Europe research and innovation program under grant agreement No. 101070238. Views and opinions expressed are however those of the author(s) only and do not necessarily reflect those of the European Union. Neither the European Union nor the granting authority can be held responsible for them.}
\thanks{P.~Jimenez, R.~Cardoso, M.~Gomes~de~Queiroz, M.~Abdalla, C.~Marchand and X.~Letartre are with Ecole Centrale de Lyon, INSA Lyon, CNRS, Universite Claude Bernard Lyon 1, CPE Lyon, 
INL, UMR5270, 69130 Ecully, France}
\thanks{F.~Pavanello is with Univ. Grenoble Alpes, Univ. Savoie Mont Blanc, CNRS, Grenoble INP, CROMA, 38000, Grenoble, France}
\thanks{M.~Abdalla is also affiliated with School of Engineering, RMIT University, Melbourne, VIC 3000, Australia}}

\maketitle

\begin{abstract}

The study of regularity in signals can be of great importance, typically in medicine to analyse electrocardiogram (ECG) or electromyography (EMG) signals, but also in climate studies, finance or security. 
In this work we focus on security primitives such as Physical Unclonable Functions (PUFs) or Pseudo-Random Number Generators (PRNGs). Such primitives must have a high level of complexity or entropy in their responses to guarantee enough security for their applications. There are several ways of assessing the complexity of their responses, especially in the binary domain. With the development of analog PUFs such as optical (photonic) PUFs, it would be useful to be able to assess their complexity in the analog domain when designing them, for example, before converting analog signals into binary. 
In this numerical study, we decided to explore the potential of the disentropy of autocorrelation as a measure of complexity for security primitives as PUFs, TRNGs or PRNGs with analog output or responses. We compare this metric to others used to assess regularities in analog signals such as Approximate Entropy (ApEn) and Fuzzy Entropy (FuzEn). We show that the disentropy of autocorrelation is able to differentiate between well-known PRNGs and non-optimised or bad PRNGs in the analog and binary domain with a better contrast than ApEn and FuzEn. Then, we show that the disentropy of autocorrelation is able to detect small patterns injected in PUFs responses. 

\end{abstract}

\begin{IEEEkeywords}
Complexity, disentropy, Pseudo Random Number Generators (PRGNs), Physical Unclonable Functions (PUFs).
\end{IEEEkeywords}

\IEEEpeerreviewmaketitle

\section{Introduction}

\IEEEPARstart{P}{hysical} Unclonable Functions (PUFs) represent a class of physical security primitives. They play the role of a physical key protecting an object that may be digital/analog data or electronic (and photonic) hardware. For some applications, they might be perceived as the electronic or photonic analogy of the biometric characteristics of a human being, such as fingerprints \cite{maes_physically_2013}, and provide interesting solutions in the context of the Internet of Things (IoT) \cite{babaei_physical_2019}.

PUF security is based on fabrication variations, i.e, variations from the designed dimensions of the structure due to the randomness of the manufacturing process. 

The sensitivity to this randomness is what confers on the PUF its properties of uniqueness and non-clonability \cite{pavanello_recent_2021}. \\
A PUF links an input information e.g., a binary string, called a ``challenge'' to a ``response'' in an ideally deterministic way forming a so-called challenge-response pair (CRP), that is not possible to be predicted in advance.
Most of the time PUFs operate with multiple challenges, to build one or more responses from the set of all possible CRPs (CRPs library). Hence, the PUF should fulfil some characteristics \cite{pavanello_recent_2021,athanas_systematic_2013}: 
\begin{itemize}
\item Physical uniqueness: As explained before, the PUF should act as a fingerprint and its responses should be mostly determined by the fabrication variations. Hence, for a same challenges different PUF instances should have different responses. 

\item Physical unclonability: Even with good equipment, an adversary should not be able to reproduce the PUF thanks to the unpredictability of fabrication variations.

\item Digital unpredictability: It should be impossible to model the behavior of the PUF numerically (using machine learning attacks for example).

\item Reliability: The PUF should be stable, i.e, for the same PUF instance, the same challenge should give the same response (within a certain error tolerance). This stability should hold over time, even in slightly different experimental conditions.
\end{itemize}
Depending on the application, some conditions have to be established on the challenge library size and on the quality of the responses. Especially in the case of a strong PUF, mostly suitable for authentication protocols, the CRP library should be very large to ensure that it would not be possible for an adversary to have access to a large portion of the CRP library in a reasonable amount of time \cite{ruhrmair_foundations_2009}. Furthermore, the responses should be complex enough so that responses potentially harvested by an attacker cannot be used to train an algorithm able to 
predict the rest of the CRPs and to model the PUF behavior.\\
Consequently, the responses of a PUF must not have patterns that are repeated over all instances, within each individual response of the same instance and between the responses of the same instance (if the responses are made of multiple bits). As a result, we should not detect patterns in the responses taken individually, nor in the concatenation of responses coming from the same PUF instance. Moreover, no patterns should be observed from the concatenation of responses of different PUF instances for a same challenge \cite{shiozaki_entropy_2020}.
This argument applies also to cryptographic primitives for random number generators (RNGs) in the analog domain which require the absence of repeating patterns.

There is an entire zoo of PUFs, most of which are electronic ones and operate in the binary domain as the Arbiter, SRAM, or Butterfly PUFs \cite{al-haidary_physically_2019}. Most of these electronic PUFs have responses of one bit for one given challenge. A multi-bits sequence is therefore built by concatenating the one-bit responses of the PUF for different challenges. In that case there is no single unambiguous way to assemble its responses into a sequence.
For these PUFs, the NIST test suite \cite{rukhin_statistical_2010}, especially suited for the analysis of binary data, can be used to evaluate some of the statistical properties of the responses \cite{marchand_implementation_2018}. However, PUFs can also operate in the analog domain, especially in some recent implementations of optical and photonic PUFs \cite{pavanello_recent_2021}. For some of these PUFs one challenge give a ordered response of multiple bits \cite{bosworth_unclonable_2020,tarik_scalable_2022,pappu_physical_2002,ruhrmair_optical_2013}.

Nevertheless, there is an important question to address when considering these PUFs that operate in the analog domain. 
The impact of fabrication variations affect some physical parameters, but PUFs output signals may present the need for a conversion from the analog to the binary domain to obtain responses that can be further used at a system level \cite{bosworth_unclonable_2020,tarik_scalable_2022,pappu_physical_2002,ruhrmair_optical_2013}. 
The quality of such responses will be impacted by the chosen conversion scheme: a different scheme will lead to different responses and probably different responses quality. Here by responses quality we refer to the responses complexity and entropy. Besides, post-processing may be needed to improve the responses quality. For example, in the first optical PUF by Pappu and co-workers in 2002 \cite{pappu_physical_2002}, the optical output of the PUF is a speckle pattern recorded on a $320\times240$ pixel camera that passes through a threshold filter to obtain an output in the binary domain, and finally through a Gabor transform to obtain a 2400-bit key. As discussed in \cite{ruhrmair_optical_2013}, the entropy of the output response strongly depends on this transformation. A Gabor transformed image has regularities (zebra-stripes) leading to important patterns in the responses and hence to low entropy. Therefore, the Gabor transform may remove parts of the complexity contained in the optical speckle pattern.

This discussion can be also applied to silicon-photonic PUFs. For example, in \cite{bosworth_unclonable_2020} the authors convert their analog signal in the binary domain using a suite of Hadamard matrices. They also apply different transformations on the obtained binary signals to improve their quality (equalization, grey code, and XOR).

In those cases, we observe that the final entropic quality of a PUF is in fact the result of a combination of several steps:

1) The complexity of the physical device, its sensitivity to fabrication variations and its randomness properties.

2) The quality of the analog to binary signal conversion i.e., if the conversion retains the entropy contained in the analog signal.

3) The post-processing step to increase the complexity or stability of the PUF.

Hence, while designing a PUF, it would be ideal for a researcher to be able to work independently on each of these steps. However, by using common benchmarks like the NIST test suite, it is impossible to evaluate the PUF physical quality since it requires a binary conversion.

This is also a problem when using TRNGs that operate in the analog domain, such as photonic TRNGs \cite{argyris_implementation_2010,ugajin_real-time_2017}, and it is important to extract as much information as possible about the analog signal from the TNRG if we want to analyze the impact of the binary conversion for instance.

Our goal is therefore to find a way of evaluating the physical quality of a PUF design or a TRNG by looking at the complexity of their outputs in the analog domain directly. In that case, the Shannon entropy \cite{shannon_mathematical_1948} used for binary data cannot be applied. Since one of the basic requirements is the absence of repeating patterns within or between responses/outputs, metrics based on the autocorrelation function may be ideal candidates. Therefore, in this paper, we propose to evaluate the performance of the disentropy, a metric developed by R.~V.~Ramos for quantum applications \cite{ramos_quantum_2020,castro_enhancing_2022}, but also to obtain a score based on the autocorrelation function \cite{castro_enhancing_2022, ramos_estimation_2021}. 
We will first describe the disentropy of the autocorrelation mathematically, then we will introduce other metrics that are used to evaluate the complexity of analog signals like the Approximate Entropy also used in cryprography and security. Finally, we will compare and evaluate the performance of these metrics on PRNGs and PUFs responses (analog but also binary) to assess whether the disentropy of autocorrelation is a good candidate or not to evaluate the quality of security primitives. For testing the metrics, rather than TRNG, we decided to test the metrics on common deterministic PRNGs whose random properties are easy to modify.

\section{The Disentropy of the Autocorrelation Function}

\subsection{The autocorrelation function}
The autocorrelation function measures the similarity between a function or a signal and a delayed version of itself with delay $\tau$. Castro et al. \cite{castro_enhancing_2022} and Ramos \cite{ramos_estimation_2021} claim that this function can be used to measure some randomness present in the signal. \\
For discrete signals, the autocorrelation function is given by:
\begin{equation}
r_k= \frac{c_k}{\sigma_0^2}=\frac{\frac{1}{N} \sum_{t=1}^{N-k}\left(s_t-\bar{s}\right)\left(s_{t+k}-\bar{s}\right)}{\sigma_0^2}
    \label{autocorrD}
\end{equation}
With $k$ the considered lag (delay), $s_t$ the discrete signal value at time $t$, $\bar{s}$ the mean value of $s_t$ and $\sigma_0^2$ its sample variance. It is important to note that the autocorrelation function can be either positive or negative depending on whether two events are correlated or anti-correlated.
\\As explained by Ramos in \cite{ramos_estimation_2021} the goal is to obtain a score from the function, hence, we need to map the function to a scalar. An intuitive way to obtain it would be to use the Shannon entropy $H$ defined as \cite{shannon_mathematical_1948}:

\begin{equation}
H = -\sum^{n}_{i=1}p_i\log(p_i)
\label{shannon}
\end{equation}
Eq.~\ref{shannon} is expressed in terms of a discrete set of probabilities $\{p_i\}$.

However, since the autocorrelation function can have negative values, it cannot be associated to a probability distribution. Moreover, the logarithm would not be defined on negative values so the classical definition of entropy as Shannon entropy cannot be used to map the autocorrelation function to a scalar. 

\subsection{The construction of disentropy}
In 1988 Tsallis generalised the definition of entropy for multifractals systems \cite{tsallis_possible_1988} with $q \in \mathbb R$ as:
\begin{equation}
    S_q=k\frac{1-\sum_i p_i^q}{q-1}
    \label{entropy4}
\end{equation}

Eq.~\ref{entropy4} tends to the Boltzmann-Gibbs entropy in the limit of $q\rightarrow1$.
Then, in 1994 he proposed a new way of interpreting the experimental measurements as $q-$expectation values \cite{tsallis_what_1994}. In this work he defines the generalised logarithmic function as:
\begin{equation}
    \ln_q(x)=\frac{x^{(1-q)}-1}{1-q} \ \forall x\in \mathbb R^+\textnormal{ and }q\neq1
\label{entropy5}
\end{equation}
This logarithm tends to the natural logarithm when $q\rightarrow1$ and a generalised exponential function can be attributed to it:
\begin{equation}
e_q^x=(1+(1-q)x)^{\frac{1}{1-q}} \ \forall x/(1+(1-q)x\ge 0)\textnormal{ and }q\neq1
\label{entropy6}
\end{equation}
With this new logarithm definition (see Eq.~\ref{entropy5}), it is possible to express the Tsallis $q$-entropy $S_q$ as in Eq.~\ref{Sq}.
\begin{equation}
    S_q=-k\sum_{i}p_i^q\ln_q(p_i)
\label{Sq}
\end{equation}

Then, da Silva and Ramos introduced the Lambert function $W(z)$ \cite{ramos_estimation_2021,da_silva_lamberttsallis_2019} to further extend the meaning of this definition of entropy. $W(z)$ is obtained by solving:
\begin{equation}
    W(z)e^{W(z)} = z
\label{lambert}
\end{equation}
This equation has an infinite number of solutions, but only two branches give real values for $z \in \mathbb R$. 
By taking the logarithm of Eq.~\ref{lambert}, and defining $z=p_i$ it is possible to obtain a definition of entropy as:

\begin{equation}
    S=-\sum_{i}p_i\ln(W(p_i))-\sum_{i}p_iW(p_i)
\label{BG_entropy}
\end{equation}    
The term $\sum_{i}p_iW(p_i)$ is called the disentropy. It is minimal when the entropy is maximal and vice-versa. It is important to note here that the disentropy does not contain a logarithm in its expression.

Next, this entropy is generalised using the Tsallis generalised exponential and Lambert-Tsallis $W_q$ function solution of:
\begin{equation}
    W_q(z)e_q^{W_q(z)}=z
\label{BG_gener_entropy}
\end{equation} 

By taking the Tsallis $q$-logarithm of Eq.~\ref{BG_gener_entropy}:
\begin{equation}
%\ln_q(z)=W_q(z)+\ln_q(W_q(z)) + (1-q)W_q(z)\ln_q(W_q(z))
\ln_q(z)=W_q(z) + ((1-q)W_q(z) + 1)\ln_q(W_q(z))
\label{BG_gener_entropy2}
\end{equation}

At last, a probability $p_i$ is inserted in Eq.~\ref{BG_gener_entropy} with $z=p_i$ and by using the Tsallis $q$-entropy of Eq.~\ref{Sq} neglecting $k$, one obtains \cite{castro_enhancing_2022, ramos_estimation_2021}:
    \begin{multline}
    S_q=-\sum_i p_i^q W_q\left(p_i\right)-\sum_i p_i^q \ln_q \left[W_q\left(p_i\right)\right] \\
    -(1-q)\sum_i p_i^q W_q(p_i)\ln_q \left[W_q\left(p_i\right)\right]
    \end{multline}
where the term:
 \begin{equation}
D_q=\sum_i p_i^q W_q\left(p_i\right)
\label{Dq_discret}
 \end{equation}
 
is the Tsallis q-disentropy.
Recall that the autocorrelation function can take values ranging from $-1$ to $1$. Therefore, a function $W_q(r_k)$ defined on this interval shall be found. The $W_2(z)$ function below is defined on the interval of values taken by the autocorrelation function:
\begin{equation}
W_2(z)=\frac{z}{z+1} , \ z>-1
\end{equation}
So, by replacing $z=p_i$ in Eq.~\ref{Dq_discret} one obtains a value for the disentropy in the continuous and discrete probability distributions:

\begin{equation}
D_2=\sum_{i} \frac{p_i^3}{p_i+1}
\end{equation}

Ramos has shown that this metric can be defined for functions that are not probability distributions. For example in the case of the autocorrelation function in \cite{castro_enhancing_2022, ramos_estimation_2021} and for the Wigner function in quantum mechanics in \cite{ramos_disentropy_2019}. Therefore, in the case of autocorrelation function, they obtain the disentropy given by Eq.~(\ref{D2Descrete}).

\begin{equation}
    D_2=\sum_{k=0}^{N-1} \frac{r_k^3}{r_k+1}
    \label{D2Descrete}
\end{equation}

With this metric it is possible to have a defined scalar value for the complexity of a signal using the autocorrelation function. A large positive or negative disentropy shows the presence of correlations (or anti-correlations) within the signal. Its ideal value is $D_2=0.5$ obtained if $r_0=1$ and $r_k=0$ $\forall k\neq0$.
Note that the disentropy is not defined for $r_k=-1$ i.e, it is not defined in the case of perfect anti-correlation. In that case, the metric diverges.

\section{Approximate entropy and Fuzzy entropy}

In our study of complexity measures for analog signals based on the disentropy, we need to compare its performances with those of other metrics. Below, we will discuss and compare two of the most used metrics for assessing the complexity of signals.

\subsection{The Approximate Entropy}
\label{SectionApEn}
As mentioned in the introduction, the NIST test suite for random and pseudo-rando generators generating binary signals \cite{rukhin_statistical_2010} is commonly used in cryptography and security to evaluate the quality of security primitives \cite{cherkaoui_design_2016,garcia-bosque_chaos-based_2019}. In this test suite the Approximate Entropy (ApEn) is used to measure the complexity and detect the presence of repeating patterns in binary signals. The ApEn has actually been developed by Steven M.~Pincus in 1991 \cite{pincus_approximate_1991} for any kind of vector in $\mathbb{R}^N$. This metric is currently used in medicine \cite{holzinger_applying_2012,udhayakumar_effect_2016}, but also in other domains such as climate studies \cite{delgado-bonal_analyzing_2020} or finance \cite{delgado-bonal_quantifying_2019}. The full algorithm of ApEn can be found in \cite{pincus_approximate_1991,delgado-bonal_approximate_2019} and summarized in Algorithm \ref{ApEnAlgo}. This algorithm has three parameters: $m\in \mathbb{N}^*$ called the embedding dimension, $r\in \mathbb{R}^+$ called the noise filter or scaling parameter, and $N$ the number of samples. The ApEn takes patterns of $m$ points in the signal, then identifies other patterns that are similar across the signal, and determines which of these patterns remains similar for the following $m+1$ points. In more mathematical terms, ApEn is based on the conditional probability that a signal that repeated itself for $m$ points will repeat itself for $m+1$ points \cite{chen_characterization_2007}.
 
 \begin{algorithm}[H]
\caption{The Approximate Entropy algorithm.}\label{ApEnAlgo}
Let $s\in\mathbb{R}^N$ be a time series of length $N$, and $n=N-m+1$.
 \\Define $\textbf{x}\in\mathbb{R}^m$ as:
 \\$\textbf{x}(i)=[s(i), s(i+1), ..., s(i+m-1)]$ $\forall i \in \mathbb{N}^*$ and $i\leq n$
 \\Then compute:
 \begin{equation*}
     C_i^m(r)=\frac{\textnormal{Number of values $j$ such that} \ d[\textbf{x}(i),\textbf{x}(j)]\leq r}{n}
 \end{equation*}
 With $d$ a metric comparing two vectors:
 \begin{equation*}
     d[\textbf{x}(i),\textbf{x}(j)] = \max_{k=1,...,m} (|s(i+k-1)-s(j+k-1)|)
 \end{equation*}

Next, compute $\phi^m$:
\begin{equation*}
    \phi^m(r)=\frac{1}{n} \sum_{i=1}^n \log \left(C_i^m(r)\right)
\end{equation*}
The ApEn is then defined by:
\begin{equation*}
    \textnormal{ApEn}(m,r,N)=\phi^m(r)-\phi^{m+1}(r)
\end{equation*}
\end{algorithm}
In the computation of $C_i^m(r)$, we see that for similarity the algorithm compares blocks within the resolution $r$ based on the Heaviside function: if $d[\textbf{x}(i),\textbf{x}(j)]\leq r$, then the patterns are considered similar. Note that $r$ is usually a function of the standard deviation of $s$ \cite{pincus_approximate_1991,delgado-bonal_approximate_2019}.

\subsection{The Fuzzy Entropy}
The ApEn can be biased and may indicate more similarities than contained in the series. It can also be inconsistent and sensitive to a change in $r$, and depends on the length $N$ of the series \cite{delgado-bonal_approximate_2019,chen_characterization_2007,richman_physiological_2000}. Therefore, other metrics emerged, for example the Sample Entropy (SampEn) developed by Richman and Moorman in 2000 \cite{richman_physiological_2000} fixes some of ApEn problems. In fact, they have shown that SampEn does not depend on the series length if $N$ is big enough, and is less biased. This makes SampEn interesting, but its results can still be untrustworthy for small $N$ \cite{chen_characterization_2007,azami_fuzzy_2019}. This problem is mainly linked to the fact that both SampEn and ApEn use a Heaviside function as a two state classifier for the blocks similarity. In reality, this frontier is blurry and it is not easy to determine whether a pattern belongs to one class or to the other \cite{chen_characterization_2007,azami_fuzzy_2019}. The Fuzzy Entropy (FuzEn) has been introduced \cite{chen_characterization_2007} to overcome these problems by using the fuzzy sets theory developed by Zadeh in 1965 \cite{zadeh_fuzzy_1965}. In this paper, Zadeh introduced the idea of fuzzy sets as a ``class with a continuum of grades of membership'' with a ``membership function'' $f_A(x)$ associating every object $x$ of a space $X$ to a real number in $[0,1]$. $f_A(x)$ represents this ``grade of membership'' of $x$ in $A$. The closer $f_A(x)$ is to $1$, ``the higher the grade of membership of $x$ in $A$'' \cite{zadeh_fuzzy_1965}.\\
In FuzEn algorithm, this membership function $f_A$ is therefore used to replace the Heaviside function in ApEn algorithm.

In \cite{chen_characterization_2007} Chen et al. use a family of exponential functions as membership functions. However, other membership functions can be used such as triangular, Z-shaped, constant-Gaussian as presented in \cite{azami_fuzzy_2019}. They are all functions of $r$ and of a defined distance metric comparing two vectors, as in Algorithm 1.

In our study, we will compare the results obtained with the disentropy of the autocorrelation to the results obtained with ApEn because of its use in cryptography for binary signals and FuzEn to dispose of ApEn defaults. These metrics will be first tested on different PRNGs of good and poor quality.
All metrics will be used to find regularities and patterns in their outputs; ApEn and FuzEn by looking for repeating patterns of size $m$ and $m+1$ along the signal and the disentropy of autocorrelation by comparing the signal with a delayed version of itself using autocorrelation.\\ If a signal does exhibit clear repeating patterns ApEn and FuzEn would have a value close to $0$, on the other side this signal would obtain a high $|D_2|$ value. 

\section{Some PRNGs}

\subsection{Linear congruential generator (LCG)}
This generator produces a sequence of pseudo-randomised numbers based on linear recursions given by  \cite{park_random_1988,bhattacharjee_search_2022}:. 
\begin{equation}
    x_{k+1} = ax_k+c\ \textnormal{mod}\ M 
    \label{CongruentialRNG}
\end{equation}
$x_k$ being the sequence with $k\in\mathbb{N}$, $M\in \mathbb{N}^*$ is the modulo, $x_0$ is the seed, $a$ is the multiplier and $c$ the increment, such that $x_0$, $a$, $c\ \in\mathbb{Z}_M$ \cite{bhattacharjee_search_2022}.
This generator is a common and old method to make a PRNG, the linear method with $c = 0$ has been developed in 1951 by D.H. Lehmer \cite{lehmer_mathematical_1951}. %and the linear congruential generator in 1958 by W. E. Thomson and A. Rotenberg \cite{Thomson}.

In this study, we use Lehmer generators and linear congruential generators with parameters shown in Table~\ref{tabPRNG}.
To obtain optimised PRNGs of good quality with a period of $M$, the parameters $M$, $a$, $c$ and $x_0$ must meet certain conditions \cite{park_random_1988,bhattacharjee_search_2022}. Among the requirements, in the case where $M$ is a multiple of $4$, $a-1$ should also be multiple of $4$; further, when $c=0$, the seed $x_0$ should be a co-prime of $M$. Some parameters in Table~\ref{tabPRNG} have been chosen to match the parameters of commonly used random functions as the C++11 \texttt{minstd\_rand} function or the GNU C Library \texttt{rand} function \cite{bhattacharjee_search_2022}. Since $x_0$ should be co-prime of $M$ when $c=0$, we chose $x_0=1$.

\begin{figure*}[t]
   \hspace{-0.2cm}
    \includegraphics[width=1\textwidth]{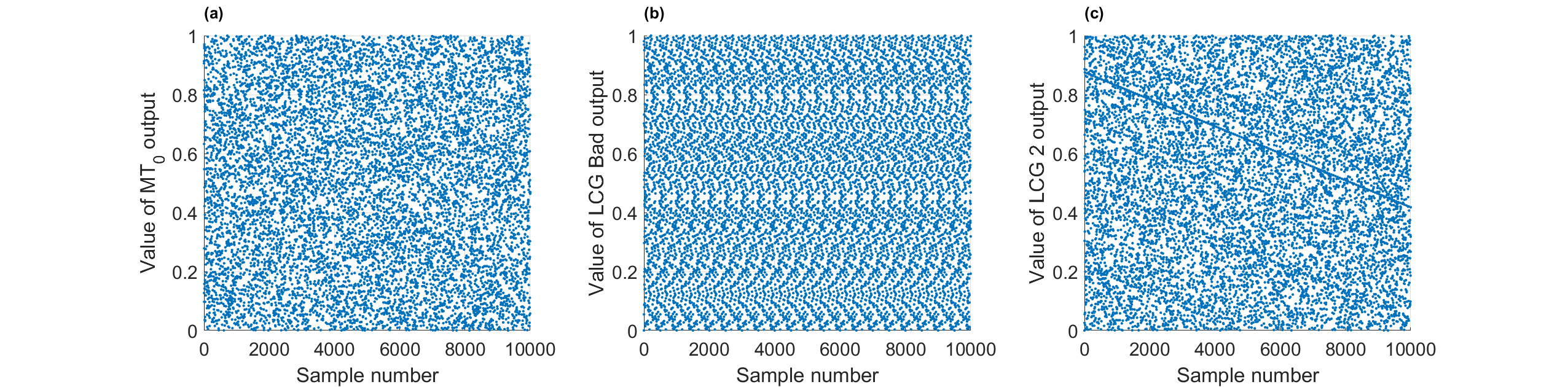}
    %\vspace{-1mm}
    \caption{Examples of normalised PRNGs output for 10000 samples (a) MT$_0$ (b) LCG Bad (c) LCG 2. Before normalisation, MT$_0$ output have been generated using the MATLAB\texttrademark \ \texttt{rand} function, LCG Bad and LCG 2 with Eq.~\ref{CongruentialRNG} equation and parameters in Tab.~\ref{tabPRNG}.}
    \label{OutputPRNG3parts}
\end{figure*}

Using Eq.~\ref{CongruentialRNG} to generate PRNGs allows us to create non-optimised ones or PRNGs exhibiting periodic patterns by choosing parameters $a$, $M$, and $c$ that do not meet the conditions required to create an optimized PRNG. These PRNGs presented in Table~\ref{tabPRNG} (LCG Bad, LCG 1, 2, 3, 4) will therefore be compared to the optimised ones using the metrics presented earlier. The parameters of LCG 1, 2, 3 and 4 have been chosen in order to observe patterns in their output visually, while still exhibiting randomness as represented on Fig.~\ref{OutputPRNG3parts}c. LCG 1, 2 and 3 have the same $a$ and $M$ but different $c$ while LCG 3 and 4 only have a different $M$.

The parameters of LCG Bad have been chosen to create a bad PRNG exhibiting periodicity as illustrated on Fig.~\ref{OutputPRNG3parts}b.\\
In this study, the output of all LCGs will be normalised between $0$ and $1$.

\begin{table} [H]
\caption{Parameters of Lehmer and linear congruential generators.} \label{tabPRNG}
\centering
\renewcommand{\arraystretch}{1.3}
\begin{tabular}{|c|c|c|c|}
\hline PRNG                                                    & $M$        & $a$     &      $c$     \\
\hline \begin{tabular}[c]{@{}l@{}}Minimal standard\\ generator \cite{park_random_1988}\end{tabular} & $2^{31}-1$ & 16807        & $0$      \\
\hline \begin{tabular}[c]{@{}l@{}}C++11\\ \texttt{minstd\_rand} \end{tabular}         & $2^{31}-1$  & $48271$      & $0$   \\
\hline \begin{tabular}[c]{@{}l@{}}GNU C Library\\ \texttt{rand} \end{tabular}         & $2^{31}$    & $1103515245$ & $12345$ \\ 

\hline LCG Bad         & $5000$    & $17$ & $256$ \\

\hline LCG 1       & $2^{20}$    & $1487$ & $25436$ \\

\hline LCG 2       & $2^{20}$    & $1487$ & $25236$ \\

\hline LCG 3       & $2^{20}$    & $1487$ & $25336$ \\

\hline LCG 4     & $2^{19}$    & $1487$ & $25336$ \\

\hline
\end{tabular}
\end{table}

\subsection{Mersenne-Twister PRNG}

The Mersenne-Twister (MT) algorithm \cite{matsumoto_mersenne_1998} is a common tool to generate sequences of random numbers. For example, it is used in python \texttt{random} module \cite{PythonMT} and MATLAB\texttrademark  \ \texttt{rand} function \cite{MatlabMT}. Furthermore, the MT algorithm can be used as a base for cryptographic cyphers, for example CryptMT \cite{matsumoto_cryptographic_2005}, or image watermarking techniques \cite{prasad_hybrid_2016}.
MT uses the twisted generalized feedback back shift register (TGFSR) algorithm developed in \cite{matsumoto_twisted_1992}. The MT19937 algorithm described in \cite{matsumoto_mersenne_1998} has a remarkably large prime period $M=2^{19937}-1$ making it a good PRNG.\\
In this work, the MT19937 algorithm in MATLAB\texttrademark \ \texttt{rand} function will be used with seed $0$ and a randomly picked seed i.e., $S=1773456103$, two PRNGs which will be called MT$_0$ and MT$_S$, respectively. Now that we have defined which PRNGs are going to be used in this study, we can apply to them the various metrics discussed so far.

\section{PRNGs Results}

\subsection{Tests on analog signals}

\label{sectionAnalogPRNG}
For the ApEn metric, we utilised the MATLAB\texttrademark \ \texttt{approximateEntropy} function \cite{MatlabApEn}. It is recommended to use the ApEn algorithm with $m=2$ or $m=3$ as well as $r$ between $0.1\sigma_0$ and $0.2\sigma_0$ \cite{pincus_approximate_1991,delgado-bonal_approximate_2019}. Therefore, we took $m=2$ and $m=3$, and the \texttt{approximateEntropy} function uses $r = 0.2\sigma_0$.

For the FuzEn metric, we used the algorithm provided in \cite{azami_fuzzy_2019} with a Gaussian membership function as recommended for long signals for a faster computation time. We also took $m = 2$ and $m = 3$ as well as and $r = 0.1253\sigma_0$ as recommended in the algorithm. 

A sweep in $m$ from $m = 2$ to $m = 8$ has been performed for ApEn and FuzEn, results are shown in Appendix \ref{mSweep} and confirm the choice of $m = 2$ and $m = 3$.

To compute the autocorrelation function we used the\\ MATLAB\texttrademark \ \texttt{autocorr} function \cite{MatlabAutoCorr} for each lag (delays) from $\textnormal{lag}=0$ (no delay) to $\textnormal{lag} = \textnormal{length}(s)-1$.\\ Next, the disentropy of the autocorrelation function ($D_2$) is obtained using Eq.~\ref{D2Descrete} by summing for each lag value.

Recall here that $D_2$ can be positive or negative with an ideal value at $D_2=0.5$ in the absence of patterns in the signal \cite{ramos_estimation_2021}.
Therefore, we decided to focus on $\mathcal{D}=|D_2-0.5|$  to better appreciate the variations around 0.

\begin{figure}[H]
   %\hspace{-0.5cm}
   \centering
    \includegraphics[width=0.4\textwidth]{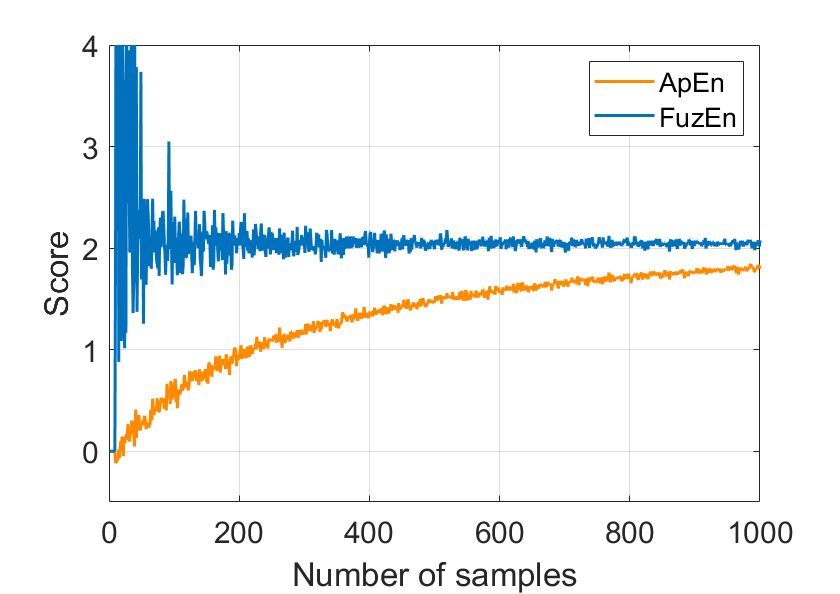}
    \caption{Convergence study of ApEn and FuzEn for MT$_0$ and $m=2$.}
    \label{FigConvergence}
\end{figure}

To know how many samples are needed to perform the analysis, we analysed how ApEn and FuzEn with $m=2$ evolve with the number of samples. Results obtained by ApEn and FuzEn applied on MT$_0$ are shown in Fig.~\ref{FigConvergence}.
With the parameters defined previously, ApEn needs more than 1000 samples to converge. The dependency of ApEn on the number of samples was expected as discussed in Section~\ref{SectionApEn}. FuzEn converges faster than ApEn, but still needs at least 1000 samples to reduce its oscillations to $\sigma_{FuzEn}\sim10^{-2}$ in order of magnitude. On the other hand, the disentropy tends to $0$ after few oscillations; for 1000 samples its oscillations are in the order of $\sigma_{\mathcal{D}}\sim10^{-3}$. It is not represented on Fig.~\ref{FigConvergence} for reasons of visibility. 

In the next studies, we chose to take a number of samples $N = 10000$ to make sure that both ApEn and FuzEn have enough samples to converge and have small oscillations. 

After 10000 samples, the oscillations for ApEn and FuzEn are in the order of $\sigma_{FuzEn}\sim\sigma_{ApEn}\sim10^{-3}$, and $\sigma_{\mathcal{D}}\sim10^{-4}$ for the the disentropy. Hence, we generated outputs of 10000 samples for each PRNG.
For example, the output of MT$_0$, LCG Bad, and LCG 2 are presented on Fig.~\ref{OutputPRNG3parts}.

 \begin{figure*}[b!]
    \hspace{-1.35cm}
        \includegraphics[width=1.07\textwidth]{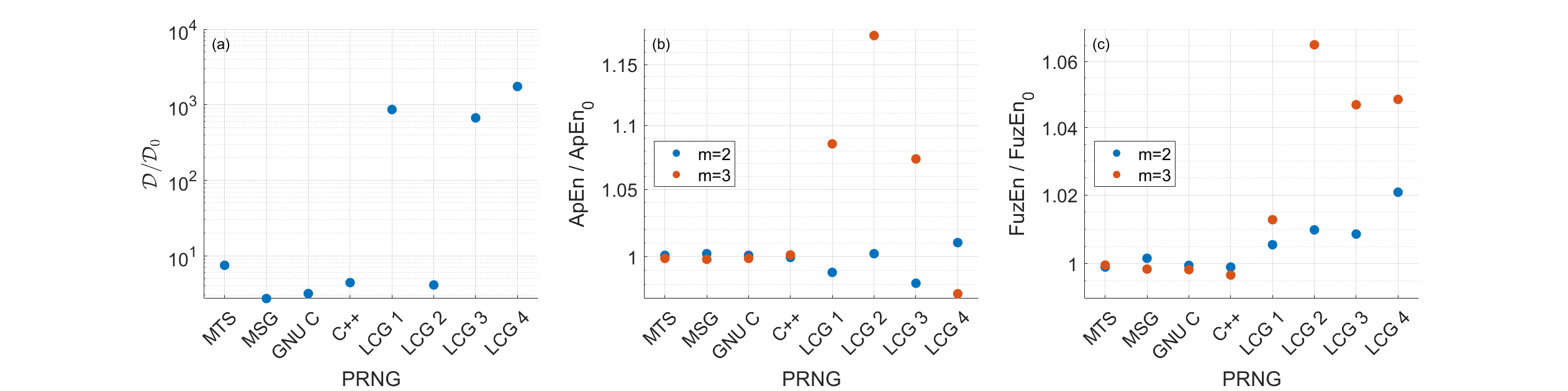}
    \caption{Score ratio of PRNGs obtained from (a) $\mathcal{D}=|D_2-0.5|$, (b) ApEn, and (c) FuzEn compared to MT$_0$ scores for analog signals of Table.~\ref{tabAnalogMT0}.}
    \label{FigAnalogResult}
\end{figure*}

We observe on Fig.~\ref{OutputPRNG3parts}a that the output of MT$_0$ does not exhibit observable patterns. On the other hand, by looking at Fig.~\ref{OutputPRNG3parts}b we see that LCG Bad has clear repeating patterns and should have bad results for all the metrics, i.e, ApEn and FuzEn close to zero and a high disentropy. As seen on Fig.~\ref{OutputPRNG3parts}c, LCG 2 does not necessarily have repeating patterns, but straight lines can be observed in its output.

Then, we decided to apply the different metrics on the PRNGs and compare their results to the scores obtained by MT$_0$ (for each respective metric) presented in Tab.~\ref{tabAnalogMT0}.

\begin{table}[h]
\centering
\caption{MT$_0$ scores for analog signals} \label{tabAnalogMT0}
\renewcommand{\arraystretch}{1.4}
\begin{tabular}{c|c|c|c|}
\hline  \multicolumn{1}{|c|}{$\mathcal{D}_0$} & & ApEn$_0$ & FuzEn$_0$ \\
\hline  \multicolumn{1}{|c|}{$4.80\cdot10^{-5}$}  & $m=2$  & $2.156$ & $2.040$ \\
\hline  &$m=3$& $1.848$ & $1.926$ \\
\cline{2-4}

\end{tabular}
\end{table}

Results are shown on Fig.~\ref{FigAnalogResult}. All metrics were clearly able to distinguish the repeating patterns of LCG Bad, but in order to have a proper visibility of the other PRNGs results, it was decided to remove its scores from Fig.~\ref{FigAnalogResult}. However, results from LCG Bad are presented in the Appendix~\ref{AnnexeLCGBad}. 

As expected, we observe that the different good or well-known PRNGs (MT$_S$, MSG, GNU C, C++) obtain good scores for all metrics, close to those of MT$_0$ with $D\sim10^{-4}$. However, only the disentropy is capable of clearly distinguishing between these known PRNGs and LCG 1, 3 and 4. FuzEn even perceives them as better PRNGs. It can be also seen that $m = 3$ obtains better contrast than $m = 2$ for ApEn and FuzEn. 

However, one observes that LCG 2 obtains a good score for the disentropy and the best one for ApEn and FuzEn with $m = 3$. This means that $\mathcal{D}$, ApEn and FuzEn are not capable of perceiving the lines observed for LCG 2 in Fig.~\ref{OutputPRNG3parts}(c). These lines are actually dotted lines with a certain periodicity; this period $p_{line}$ has an impact on the metrics scores. A test has been conducted with MT$_0$ output in which we included a diagonal line of a certain periodicity. With $m=3$, ApEn and FuzEn are not able to detect the line for $p_{line}>3$. With $m=2$ they don't detect it for $p_{line}>2$, we also tested with $m=4$ and concluded that ApEn and FuzEn would not detect the line for $p_{line}>m$. On the other side, the disentropy is not capable of detecting the line for $p_{line}>40$ while the line with the smallest $p_{line}$ in LCG2 has a period of $64$ samples.

\begin{figure}[H]
   \centering
    \includegraphics[width=0.47\textwidth]{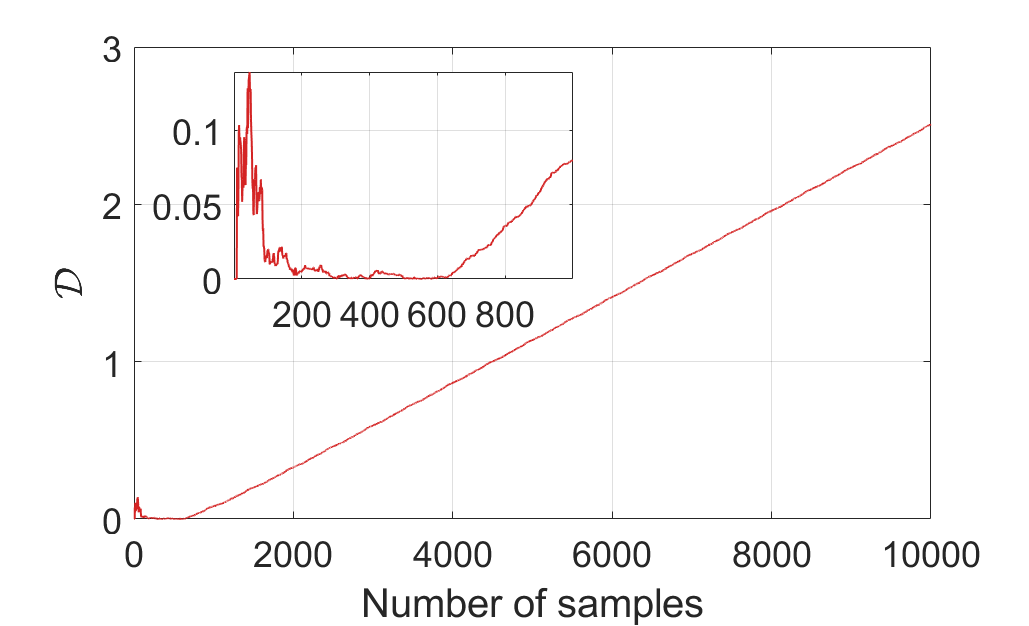}
    \caption{Evolution of LCG Bad disentropy with $N$.}
    \label{ConvergenceBadPRNG}
\end{figure}

By looking at Eq.~(\ref{D2Descrete}), it is possible to deduce that the score obtained by the disentropy will depend highly on the number of samples if the pattern repeats itself periodically. Indeed, if one defines $T$ as the period of the pattern, and $\tau_p$ a point in the periodic pattern, then $r_{\tau_p} = r_{\tau_p+T}>0$. Hence, the contribution of the periodic pattern will be added each time it is observed. This behavior has been verified with LCG Bad that repeats itself every 500 samples. Fig.~\ref{ConvergenceBadPRNG} shows that the disentropy score is small for a small number of samples i.e., below $500$. However, above $500$ samples the metric begins to analyze redundancies and the score $\mathcal{D}=|D_2-0.5|$ grows linearly with the number of samples due to the increase of autocorrelation functions $r_{500j}$ with $j\in\mathbb{N^*}$.
%\vspace{-0.2cm}

The metrics discussed thus far can also be applied to TRNGs allowing us to compare with other generators. Here we used a TRNG based on atmospheric noise \cite{haahr_randomorg_nodate}. As presented in Appendix \ref{Annexe_TRNG}, for the different metrics, the TRNG obtains results similar to MT$_0$ and MT$_S$. This means that by taking only 10000 samples, much smaller than the period of MT, the different metrics are not able to detect patterns in MT outputs and consider it similar to a TRNG.

\subsection{Tests on binary signals}
Since Eq.~(\ref{autocorrD}) is also defined for binary signals we want to test how the disentropy behaves in that case. 

\begin{figure*}[b!]
   \hspace{-1.5cm}
        \includegraphics[width=1.07\textwidth]{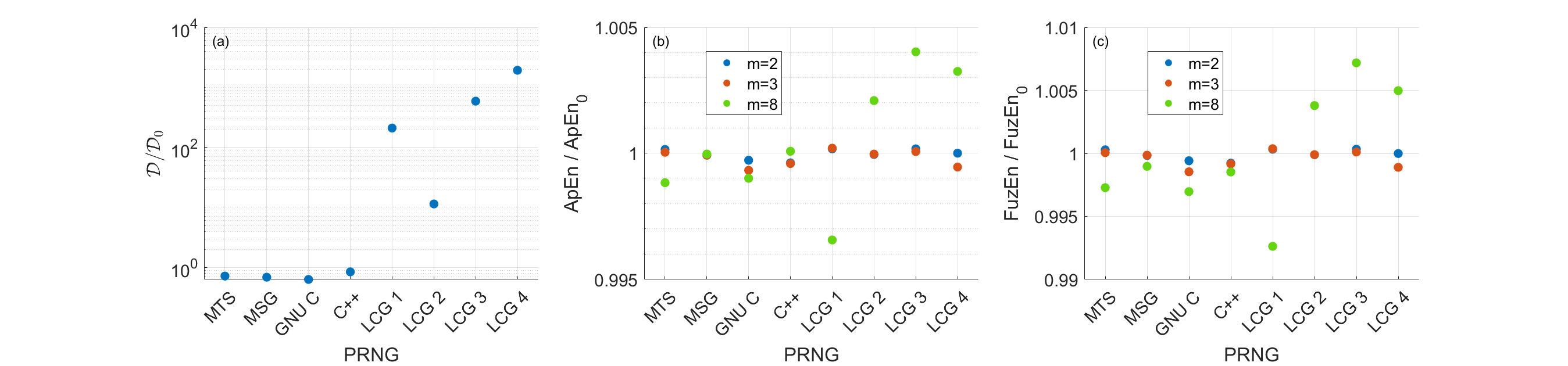}
    \caption{Score ratio of PRNGs obtained from (a) $\mathcal{D}=|D_2-0.5|$ (b) ApEn (c) FuzEn compared to MT$_0$ scores for binary signals of Table.~\ref{tabBinaireMT0}.}
    \label{FigBinaryResult}
\end{figure*}

 By placing a comparator at $0.5$ on the PRNGs output of section \ref{sectionAnalogPRNG} we converted these analog signals into binary ones. For binary signals the NIST test suite manual \cite{rukhin_statistical_2010} recommend to have $m<\lfloor \log_2(N)\rfloor-5$. With $N=10000$ this condition becomes $m<8$, hence we decided to compute the entropies with $m\in[2,8]$ with the last point $m=8$ included. We decided to do the same analysis as in Fig.~\ref{FigAnalogResult} by comparing the PRNG scores to the MT$_0$ scores shown in Tab.~\ref{tabBinaireMT0}. 
 
 \begin{table}[H]
\centering
\caption{MT$_0$ scores for binary signals.} \label{tabBinaireMT0}
\renewcommand{\arraystretch}{1.4}
\begin{tabular}{c|c|c|c|}
\hline \multicolumn{1}{|c|}{$\mathcal{D}_0$} &  & ApEn$_0$ & FuzEn$_0$ \\
\hline \multicolumn{1}{|c|}{$2.63\cdot10^{-4}$} & $m=2$  & $0.6930$ & $0.6932$ \\
\hline & $m=3$ & $0.6929$ & $0.6935$ \\
\cline{2-4} & $m=8$ & $0.6798$ & $0.6931$ \\

\cline{2-4}

\end{tabular}
\end{table}
First, one observes that ApEn$_0$ and FuzEn$_0$ scores decrease from values close to $2$ in Tab.~\ref{tabAnalogMT0} to values close to $\log(2)\approx0.6931$. In fact, the highest ApEn value for binary signals is known to be equal to $\log(2)$ \cite{delgado-bonal_approximate_2019}.

As in the analog case, all metrics were able to discriminate LCG Bad; its scores are presented in the Appendix \ref{AnnexeLCGBad}. 

Results for the other PRNGs are presented in Fig.~\ref{FigBinaryResult} where we observe that only the disentropy is capable to distinguish between the well-known PRNGs and LCG 1, 2, 3 and 4. In Fig.~\ref{FigBinaryResult}(a) and Fig.~\ref{FigBinaryResult}(b) we decided to represent $m=2$ and $m=3$ because they were the inner dimensions used in Section~\ref{sectionAnalogPRNG} and $m=8$ because it exhibited the highest contrast between the PRNGs. The well-known PRNGs still obtain results close to $\mathcal{D}_0\sim10^{-4}$. On the other hand, ApEn and FuzEn are only capable to discriminate LCG Bad and LCG 1. By increasing $m$ further, both are more and more discriminated, but all the other PRNGs obtain scores around $\log(2)$.
 
 It is possible to obtain a $p$-value from ApEn in the NIST test suite to discriminate between random and non-random binary series; if $p>0.01$ the series is considered random \cite{rukhin_statistical_2010}. Therefore, the ApEn algorithm from \cite{ang_stevenangrandomness_testsuite_2024} has been used. 
 
 This algorithm obtains results for $m=2, 3$ and $8$ similar to the MATLAB\texttrademark \ \texttt{approximateEntropy} function. In terms of $p$-values, only LCG Bad is considered as non-random except for $m=8$ where LCG 1 obtains $p=0.0096$ and is therefore near the limit.\\
Here, the disentropy of autocorrelation outperformed ApEn and FuzEn and, this time, it is capable of discriminating LCG 2 from the other good PRNGs with one order of magnitude difference.

\subsection{Test on multi-level analog signals}

In this section we want to investigate how the different metrics are impacted by the number of levels, as we saw previously, ApEn and FuzEn have their results going from approximately $2$ for MT$_0$ analog signal and $\log(2)$ for binary signals. Here, we will treat the different signals as if they were analog signal oscillating between multiple levels. To do so, we generate an analog signal of 10000 samples with MT$_0$ and cut the interval [0,1] into multiple levels. The numbers generated with MT$_0$ are assigned to a level depending on their value. The analysis is performed from two levels
$\{0,1\}$ to ten levels 
$\{0, 0.1111, 0.2222, 0.3333, 0.4444, 0.5556, 0.6667, 0.7778,\\ 0.8889, 1\}$. Furthermore in order to take into account the fact that an analog signal can oscillate around multiple levels, we add a small Gaussian noise (MT) around the levels using the MATLAB\texttrademark \ \texttt{normrnd} function with $\mu=0$ and a $\sigma\in[0,0.1]$. 

\begin{figure*}[t!]
\centering
        \includegraphics[width=\textwidth]{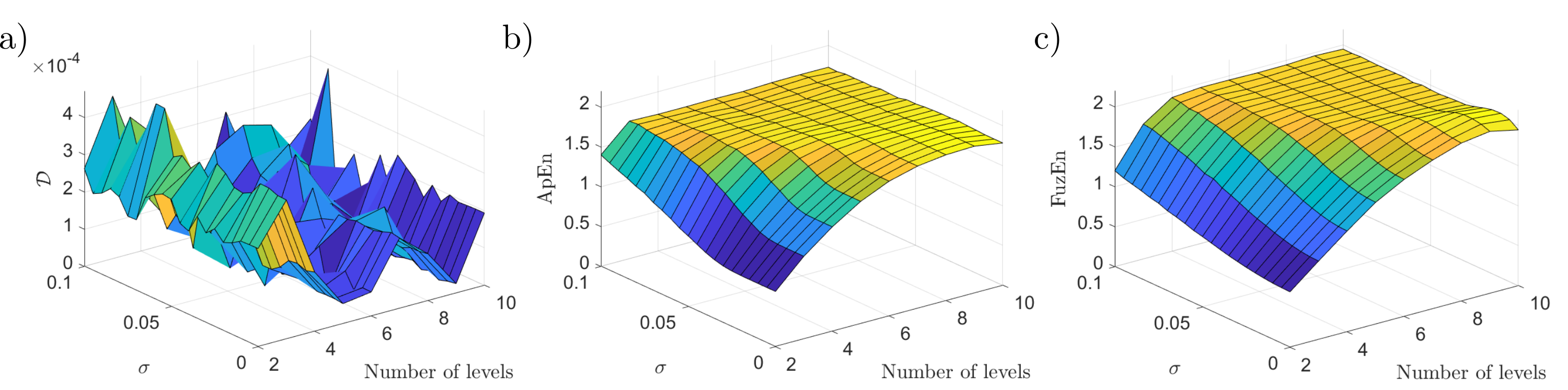}
    \caption{Evolution of a) ApEn$(m=3)$ b) FuzEn$(m=3)$ c) $\mathcal{D}$ with the number of levels of a multi-level signal generated MT$_0$. A Gaussian noise with standard deviation $\sigma$ has been added to each level.}
    \label{ApEn_FuzEn_D_levels}
\end{figure*}
For ApEn, FuzEn we used $m=3$. As shown on Fig.~\ref{ApEn_FuzEn_D_levels}a) and b), ApEn and FuzEn start around $\log(2)$ with two levels and $\sigma=0$, ApEn reaches a plateau around $1.8$ and FuzEn $1.9$ for ten levels giving results close to analog MT$_0$ results of Tab.\ref{tabAnalogMT0}. The fact of adding noise around the levels make ApEn and FuzEn perceive the signal ``more analog'' for ApEn and FuzEn. ApEn and FuzEn do not seem to be the metric to use when the frontier between analog and binary is less explicit than in the case where we clearly have '0' and '1' and in the case we have an analog signal oscillating between a significant amount of levels. On the contrary and as already observed on Fig~\ref{FigAnalogResult}(a) and Fig.~\ref{FigBinaryResult}(a), the disentropy is not highly affected by the transformation from analog to binary signal (c.f the well-known PRNGs and LCG Bad results). It is not very affected by the noise neither. As the multi-level signal is generated using MT$_0$, therefore with the same amount of regularities as MT$_0$, we expect a disentropy score close to the one of MT$_0$. The fact of adding multiple sub-levels with the Gaussian noise using MT does not affect the disentropy neither.

Hence, one should be careful while using ApEn and FuzEn on analog signals if they oscillate between two (few) levels. As mentioned in \cite{delgado-bonal_approximate_2019}, the ApEn algorithm gives a relative value allowing to compare signals using the same ``alphabet''. Therefore, depending on the application, this characteristic of ApEn and FuzEn to be sensitive to the type of signals can make a comparison between analog signals complicated. Furthermore, as seen on Tab.~\ref{Sweep_m_LCGBad} and Fig.~\ref{FigBinaryResult} the choice of $m$ depends on the type of signal (analog/binary) and tests should be performed to find the correct $m$ to apply on the signal. 

One could argue that a binary signal is less complex than an analog signal with a very large number of levels. However, for security applications where the signal has to be converted into binary data, the detection of patterns and correlations in the signal may be more important than the complexity in terms of levels and the instability of ApEn and FuzEn can be detrimental in this regard.

\section{PUFs}

In this section we initially consider series of analog signals that are not experimental data generated by PUFs. 

First, we generate the responses with the MT algorithm, then we insert patterns inside and, finally, we test them.

We generated the responses $R_i$ of $n=128$ samples as if they were responses coming from the same analog PUF instance or responses coming from different PUF instances to be converted in binary. As a reference, 100 different responses with the MATLAB\texttrademark \ \texttt{rand} function were taken, then we concatenated them to obtain a series $C_0$ of size $N=12800$. From an application point of view analyzing the concatenation of responses of different PUF instances to the same challenge could be used to detect correlations between PUF instances for example as explained in \cite{shiozaki_entropy_2020}. Detecting patterns between instances or between responses of the same instances can also give information about the PUF to an attacker. The reader may wonder how to order the concatenation of PUFs responses to extract as much information as possible, this would not be addressed in this paper but any patterns found in any responses ordering might give information about the PUF to an attacker in case they have access to PUF instances. 

As no patterns are present in the responses, the concatenation $C_0$ should obtain good scores. Besides, we generated 128-sample-long responses with \texttt{rand}, but this time we added the conditions presented in Algorithm~\ref{algo1}.

\begin{algorithm}
\begin{algorithmic}%[1]
\For{$k\in [2,n]$}
    \If{$s(k-1)\geq 0.9$} 
        \State $s(k) \gets 0.1$
    
    \ElsIf{$s(k-1)\leq 0.1$}
        \State $s(k) \gets 0.9$

    \EndIf
\EndFor
\caption{Deterministic dynamics algorithm.}\label{algo1}
\end{algorithmic}
\end{algorithm}

By doing so, we insert two patterns in the responses of the PUF with a probability of $p=0.1$ each. This simulates a PUF with a certain deterministic dynamics. These responses will be the test responses and will be compared to the reference ones for each metrics. This study has been performed 100 times to emulate 100 different PUF instances. For each instance we obtained the concatenation of their responses $C_i$ with $i \in \mathbb{N}^*$ and $i\leq 100$.

For each $C_i$ we compute the metrics for $m=1, 2$ and $3$. We decided to use $m=1$ because the patterns implied two samples. Therefore, ApEn and FuzEn might perform better with $m+1=2$ considering Algorithm~\ref{ApEnAlgo}.

\begin{table}[H]
\centering
\caption{Average metric score relative difference of $C_i$ between the test and reference responses.} \label{tabPUF0901}
\renewcommand{\arraystretch}{1.3}
\begin{tabular}{c|c|c|c|}
\hline  \multicolumn{1}{|c|}{$\mathcal{D}$} & & ApEn  & FuzEn \\
\hline  \multicolumn{1}{|c|}{$+27742\%$}  & $m=1$  & $-15\%$ & $-26\%$\\

\hline   &  $m=2$ & $-20\%$ & $-16\%$\\

\cline{2-4}  &  $m=3$  & $-12\%$ &  $-19\%$\\
\cline{2-4}

\end{tabular}
\end{table}

In Table~\ref{tabPUF0901}, we observe that all the metrics were able to differentiate between the reference and the test responses. However, for the test responses, the $\mathcal{D}$ scores increased by $\sim10^{2}$, while ApEn and FuzEn only decreased by $\sim10^{-1}$ in order of magnitude.

It is possible to analyse the responses individually by removing the concatenation step. In that case, the metrics are applied on series of 128 samples making it more difficult to differentiate between the reference responses and the test responses. The study has been performed with 500 responses to have enough data for the mean values of the metrics to converge. Between the reference and the test signals, the mean value of $\mathcal{D}$ over all responses increased by $331\%$. FuzEn($m=1$) and FuzEn($m=3$) decreased by $25\%$ and $20\%$, respectively and ApEn($m=1$) decreased by $13\%$, while ApEn($m=3$) increased by $62\%$. 

Note that ApEn($m=3$) gives a wrong result in that case, in fact its mean value should decrease as ApEn($m=1$) and FuzEn. This error is probably due to shortness of the responses that hinders ApEn to give meaningful results as the pattern is too short to be observed with $m=3$.

Now we consider 100 instances of a PUF that has a defect where some samples will always be equal. For example, one can set the value of the first sample, forcing it to be equal to 0.5. In that case, no metrics are able to differentiate between the reference and the test concatenations. If we force the first two response samples to be equal to $0.2$ and $0.1$ respectively, the [0.2, 0.1] pattern will repeat itself every 128 bits in the concatenation. 
The tests are performed with different number of responses $N_{resp}$. Intuitively, by increasing the number of responses, we increase the length of the concatenation and number of repeating patterns inside. Results are presented in Tab.~\ref{tabPUF0201}. We decided to put $0\%$ if the relative difference between the reference and test responses is lower than $1\%$ and smaller than the standard deviation of the metrics over the $100$ instances.

\begin{table}[H]
\centering
\caption{Average metric score relative difference of $C_i$ between the test responses with the [0.2, 0.1] pattern and reference responses.} \label{tabPUF0201}
\renewcommand{\arraystretch}{1.3}
\begin{tabular}{|c|c|c|c|}
\hline Metrics & $N_{resp}=100$  & $N_{resp}=200$ & $N_{resp}=500$\\
\hline $\mathcal{D}$   & $0\%$  & $+719\%$ & $+4930\%$\\

\hline ApEn($m=1$)  & $0\%$ & $-0.3\%$ & $-0.3\%$\\

\hline FuzEn($m=1$)   & $-0.5\%$ & $-0.5\%$  & $-0.5\%$ \\

\hline ApEn($m=2$)  & $0\%$ & $-0.3\%$ & $-0.3\%$\\

\hline FuzEn($m=2$)   & $-0.4\%$ & $-0.3\%$  & $-0.3\%$ \\

\hline ApEn($m=3$)  & $0\%$ & $-0.2\%$ & $-0.2\%$\\

\hline FuzEn($m=3$)   & $0\%$ & $-0.3\%$  &  $-0.3\%$\\

\hline
\end{tabular}
\end{table}
For $N_{resp}=100$, we observe that only FuzEn is capable of detecting the pattern with a very small contrast of $0.5\%$ for $m=1$ and $0.4\%$ for $m=2$. However, by increasing the number of responses, and therefore by increasing the number of patterns present in the signal, the disentropy contrast increases drastically, while the FuzEn and ApEn contrast remains very small for all $m$. 

\section{Conclusion}
In this study we were able to show that the disentropy of autocorrelation is an interesting metric to assess the complexity of security primitives outputs, both in the analog as well as in the binary domain. The disentropy is able to differentiate between well-known and optimized PRNGs and ones of lower quality with a greater contrast than ApEn and FuzEn. It can also be used to detect patterns in the analog outputs and binary responses of PUFs.\\
Furthermore, the disentropy does not need inputs whereas ApEn and FuzEn results depend highly on their input parameters such as the inner dimension $m$. In addition, ApEn and FuzEn depend on the type of the input signal $s$; with the chosen $m$ and $r$ we observed that the ideal value for ApEn and FuzEn is close to $2$ for analog signals and $\log(2)$ for binary ones or analog signals fluctuating between two levels. On the other hand, the results obtained from the disentropy do not suffer from this characteristic since it only detects the amount of correlation in signals.\\ %making it more interesting to use for signals that have to be converted into binary data. 
However, we saw that none of the metrics were able to detect the lines in LCG~2 analog output. ApEn and FuzEn failed in detecting these lines for a period $p_{line}>m$ and the disentropy for $p_{line}>40$. Also, the score given by the disentropy depends highly on the length of the signal. Therefore, we recommend using it with a comparison. We suggest to compare the score obtained by the signal under test alongside a signal of the same length generated with the MT$_0$ algorithm or a TRNG like an atmospheric noise-based TRNG \cite{haahr_randomorg_nodate}.

\appendices

\section{Influence of the inner dimension on ApEn and FuzEn scores for analog signals}
\label{mSweep}

\begin{figure}[H]
   \centering
    \includegraphics[width=0.4\textwidth]{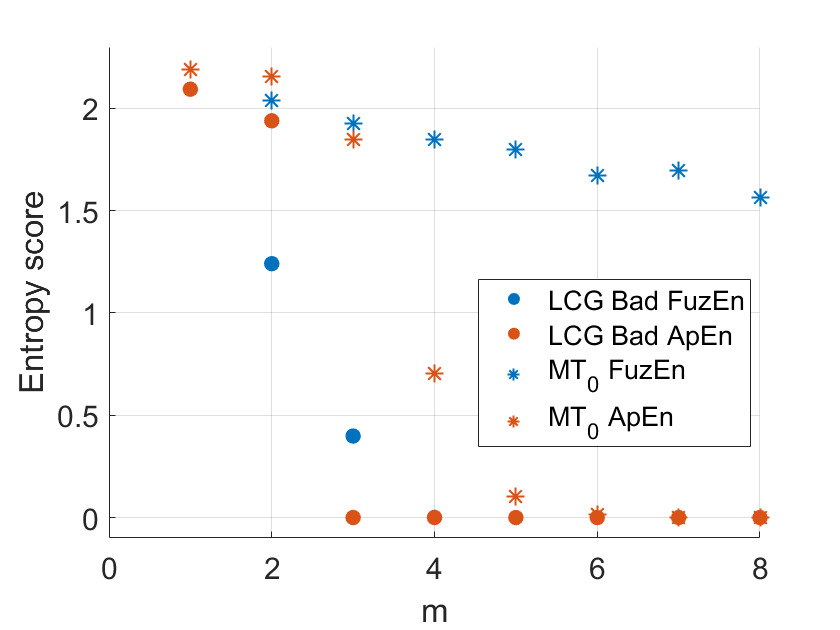}
    \vspace{-1mm}
    \caption{LCG Bad and MT$_0$ entropy score evolution with $m$ for analog signals.}
    \label{Sweep_m_LCGBad}
\end{figure}

To confirm the choice of $m=2$ and $m=3$ for ApEn and FuzEn we decided to conduct a study of the influence of $m$ in the ApEn and FuzEn scores. As shown in Fig.~\ref{Sweep_m_LCGBad}, with $m=1$ we cannot differentiate between MT$_0$ and LCG Bad. Using $m=2$ we obtain a high score for MT$_0$ as expected, but a small contrast compared to LCG Bad. $m=3$, seems to be the best candidate since it has the largest contrast between the two PRNGs while giving a high score for MT$_0$. For $m>3$, the contrast decreases and ApEn gets close to $0$ for MT$_0$.

\section{LCG Bad scores}
\label{AnnexeLCGBad}

The scores obtained by LCG Bad are hard to represent in Fig.~\ref{FigAnalogResult} and Fig.~\ref{FigBinaryResult} without losing visibility for the other PRNGs. Therefore, its results are presented in Tab.~\ref{tabAnalogLCGBad} for analog signals and in Tab.~\ref{tabBinaireLCGBad} for binary signals.

\begin{table}[h]
\centering
\caption{LCG Bad scores for analog signals.} \label{tabAnalogLCGBad}
\renewcommand{\arraystretch}{1.4}
\begin{tabular}{c|c|c|c|}
\hline \multicolumn{1}{|c|}{ $\mathcal{D}$} &  & ApEn & FuzEn \\
\hline \multicolumn{1}{|c|}{$2.51$} & $m=2$ & $1.939$ & $1.241$ \\
\hline & $m=3$ & $0$ & $0.398$ \\
\cline{2-4}

\end{tabular}
\end{table}

\begin{table}[h]
\centering
\caption{LCG Bad scores for binary signals.} \label{tabBinaireLCGBad}
\renewcommand{\arraystretch}{1.4}
\begin{tabular}{c|c|c|c|}
\hline \multicolumn{1}{|c|}{ $\mathcal{D}$} & & ApEn & FuzEn \\
\hline  \multicolumn{1}{|c|}{$2.60$} & $m=2$ & $0.692$ & $0.690$ \\
\hline & $m=3$ & $0.685$ & $0.677$ \\
\cline{2-4} & $m=8$ & $0.367$ & $0.381$ \\
\cline{2-4}
\end{tabular}
\end{table}

\section{Atmospheric noise TRNG}
\label{Annexe_TRNG}
In this section we will generate sequences from data generated from the website at Ref.~\cite{haahr_randomorg_nodate}. It uses a TRNG based on atmospheric noise. The service has been built by Dr.~Mads Haahr of the School of Computer Science and Statistics at Trinity College in Dublin and is now operated by Randomness and Integrity Services Ltd \cite{haahr_randomorg_nodate}. Atmospheric noise is generated by lightning discharges. A lightning is not composed of one plasma pulse but is a complex phenomenon composed of multiple steps and pulses generating the atmospheric noise \cite{watt_characteristics_1957,giordano_modeling_1972}. 

Since we cannot set a seed and collect the same sequence for this TRNG it is necessary to collect multiple sequences and test how they perform on average for multiple sequences generated with the TRNG. In order to compare this result with the PRNGs we decided to generate sequences of 10000 samples using the ``Random Integer Set Generator''. Each of these sets contains integers ranging from 1 to 1,000,000. The sets are collected and normalized in order to obtain a signal with 10000 samples $\in[0,1]$ before testing them with the metrics.
The number of sets/samples is limited on the website, therefore we decided to limit this statistical study to 10 sets. The reader may argue that this could be too low but it would already give a good idea of the TRNG performance. Since we want to detect patterns with 10000 samples we expect the results of the TRNG to be similar to the ones of MT$_0$ and MT$_S$ presented in Tab.~\ref{tabAnalogMT0} and Fig.~\ref{FigAnalogResult}. TRNG results are presented on Tab.~\ref{tabTRNG}.

\begin{table}[h]
\centering
\caption{TRNG score} \label{tabTRNG}
\renewcommand{\arraystretch}{1.4}
\begin{tabular}{c|c|c|c|}
\hline  \multicolumn{1}{|c|}{$\mathcal{D}$} & & ApEn & FuzEn \\
\hline  \multicolumn{1}{|c|}{$(2.10\pm 1.40)\cdot10^{-4}$}  & $m=2$  & $2.160\pm0.003$ & $2.042\pm0.003$ \\
\hline  &$m=3$& $1.846\pm0.008$ & $1.923\pm0.004$ \\
\cline{2-4}

\end{tabular}
\end{table}

These results are indeed similar to those obtained by the MT$_0$ and MT$_0$ algorithms. Therefore, we can conclude that as long as the sequence length is much smaller than the maximum period of MT, the metrics consider that MT generates a random sequences. 

\section*{Acknowledgments}

The authors would like to thank R.~V.~Ramos for the helpful and interesting discussion on disentropy and Stefano Giordano for useful feedback.

\bibliographystyle{unsrt}
\bibliography{main.bib}

\end{document}